\documentclass[aps,prb, notitlepage, twocolumn, preprintnumbers, nofootinbib,  nolongbibliography, amsmath,amssymb,superscriptaddress,floatfix,scrartcl, longbibliography]{revtex4-2}

\usepackage{graphicx} 
\usepackage{url}
\usepackage{upgreek,amssymb,multirow}
\usepackage{graphicx,color}
\usepackage[dvipsnames]{xcolor}
\usepackage{mathtools}
\usepackage{float}
\usepackage{mciteplus}
\usepackage{blindtext}
\usepackage[unicode=true,pdfusetitle,bookmarks=false,colorlinks=true,citecolor=blue,urlcolor=blue,linkcolor=red]{hyperref}

\usepackage{amsmath}
\usepackage{extarrows}
\usepackage{braket}
\usepackage{tikz}
\usepackage{tikz-cd}

\begin{document}
\title{Dissipative Dynamics and Symmetry Breaking in Bosonic Sachdev-Ye-Kitaev Lindbladian}

\author{Yifei Liu}
\affiliation{Department of Physics, Princeton University, Princeton, New Jersey 08544, USA}

\author{Anish Kulkarni} 
\affiliation{Department of Physics, Princeton University, Princeton, New Jersey 08544, USA}

\author{Shinsei Ryu}
\affiliation{Department of Physics, Princeton University, Princeton, New Jersey 08544, USA}

\begin{abstract}
We investigate a bosonic variant of the Sachdev-Ye-Kitaev (SYK) model coupled to a Lindbladian environment, focusing on the interplay between quantum many-body dynamics and dissipation.
Using the Schwinger-Keldysh path integral formalism in the large-$N$ limit,
we uncover a rich phase structure, including symmetry breaking and phase transitions. 
Our results suggest that the dissipation can partially tame the instability of the inverted potential, leading to novel steady-state phases. We also identify regimes with multiple competing saddle points and discuss potential implications for the landscape of metastable states. 
\end{abstract}

\maketitle

\section{Introduction}

The physics of open quantum systems has recently attracted growing interest.
Since coupling to the external environment is unavoidable in realistic physical
systems, an understanding of open quantum systems is of fundamental importance for quantum technologies.
Notably, dissipation is not necessarily a nuisance that destroys quantum coherence and entanglement;
rather, dissipation can be utilized and even lead to new physical phenomena
that have no analogs in closed quantum systems.
For example, engineered dissipation can be utilized to prepare a desired quantum state~\cite{Verstraete-09, Diehl-08, Diehl-11}.
Dissipation can also give rise to unique non-Hermitian topological phenomena
\cite{Ashida_2020, Bergholtz-review}.
Furthermore, open quantum systems exhibit phase transitions that cannot occur in
closed quantum systems at thermal equilibrium.
Prime recent examples include the entanglement phase transitions induced by the competition between the unitary dynamics and the quantum measurements~\cite{
Li-18, Li-19, Skinner-19, Fisher_2023}.
Despite these recent advances, the interplay of strong many-body interactions and dissipation, as well as the consequent phase transitions, has yet to be fully understood.

In this work, we propose to study many-body quantum systems coupled to an external environment, focusing on the interplay between strong interactions and dissipation. 
As a toy model to explore these questions, we consider Sachdev-Ye-Kitaev (SYK)-type models 
in the presence of dissipation, 
by using the Lindbladian quantum master equation.
The SYK model is a paradigmatic model for quantum many-body chaos, non-Fermi liquid,
and holography
-- see, e.g.,
\cite{sachdev1993,kitaev2015,sachdev2015PRX,PhysRevLett.119.216601,
chowdhury2021sachdev, brzezinska2022engineering}.
Refs.\ \cite{S__2022} and \cite{kulkarni2021} introduced
an open version of the SYK model in the context of Lindbladian dissipation.
Under the assumption of memory times much shorter than all other characteristic time scales (Markovian approximation), the time evolution equation for the system's
reduced density matrix assumes the Lindblad form
\cite{1976CMaPh..48..119L, 1976JMP....17..821G}.
Besides the Hamiltonian, describing coherent evolution, the Lindbladian includes
channels of interaction with the environment,
modeled by so-called Lindblad jump operators or dissipators, that act, e.g., as sources of
dephasing and dissipation.

As its hermitian (unitary) cousins, the SYK Lindbladian can be exactly solvable in the limit of a large number of fermion flavors,
and exhibits  many rich features,
making it an ideal playground for 
exploring
strongly interacting dissipative quantum systems.
For instance, in terms of stationary-state behavior, the model exhibits a transition between coherent and overdamped dynamics as the strength of dissipation is varied. It also displays anomalous diffusion phenomena
\cite{
Garc_a_Garc_a_2023,
PhysRevB.109.064311}.
For finite time dynamics, 
dynamical phase transitions
were found 
in the dissipative form factor
\cite{Kawabata_2023a}.
Its out-of-time-order correlators 
and operator growth
have been analyzed
\cite{Bhattacharjee_2023,
Bhattacharjee_2024,
Liu_2024,
garcíagarcía2024lyapunovexponentsignaturedissipative}.
Various other aspects, 
including symmetry properties and level statistics
\cite{Kawabata_2023c,
li2024spectralformfactorchaotic},
wormhole formation
\cite{Garc_a_Garc_a_2022,
Wang_2024},
strong to weak spontaneous symmetry breaking
detected by
coherent information 
\cite{kim2024errorthresholdsykcodes},
topological properties
\cite{Garc_a_Garc_a_2025},
and
scar states 
\cite{garcíagarcía2025lindbladmanybodyscars},
have also been studied.

In this paper, we introduce and study a bosonic variant of the SYK Lindbladian. 
While one of our main motivations for the model 
comes from the fermionic counterpart, 
there are a number of similar/related models. 
First, the non-interacting part of our model
resembles the celebrated 
Caldeira-Leggett model
\cite{Caldeira-Leggett-PRL81,
  Caldeira-Leggett-AP83,
  Leggett-review}.
On the other hand, the non-dissipative part of our model is analogous to the quantum 
spherical $p$-spin glass model 
\cite{Cugliandolo_2001, Anous_2021}.
Our model also resembles 
driven-dissipative and interacting bosonic quantum systems that have 
been studied extensively in the context of quantum optics --
see, e.g.,
\cite{Diehl_2008, 
Torre_2013,
Sieberer_2013,
Le_Boit__2013,
Hartmann_2016,
Biella_2017,
Savona_2017,
Dykman_2018,
McDonald_2018,
Rota_2019,
Roberts_2023,
mondal2025transientsteadystatechaosdissipative}
for a non-exhaustive list of recent works.
Additionally,
for other models with a similar flavor
which feature both strong and complex interactions
and coupling to the environment,
see, e.g.,  \cite{kimura2024analysisdiscretemodernhopfield,
Cho2025,
barberena2024postselectionlatticebosonsundergoing}.

Using the Schwinger-Keldysh path integral in the large-$N$ limit, we uncover rich dynamical
behaviors of the model, 
including dynamical phases that appear due to the interplay between strong correlation and dissipation.
The different dynamical phases are characterized by the number of saddle point solutions and pattern of symmetry breaking -- these features do not exist in the fermionic counterpart.

\section{Model}

The model we consider in this paper is
defined in terms of $N$ copies of 
the Heisenberg algebras, 
$[x_i, p_i]= i$, $i=1,\cdots, N$.
The Hamiltonian is given by 
a simple harmonic oscillator potential
and an SYK-like coupling, 
\begin{align}
\label{eq: model}
    H = 
    \sum_i
    \frac{p_i^2}{2m}
    + v \sum_i  x^2_i
    +
    \sum_{i_1,\cdots, i_q} J_{i_1\cdots i_q}x_{i_1} \cdots x_{i_q},
\end{align}
where $J_{i_1 \cdots i_q}$ is a real Gaussian random variable with zero mean and variance
\begin{align}
\label{bosonic SYK}
	\langle{(J_{i_1 \cdots i_q})^2}\rangle 
    = \frac{J^2}{2N^{q-1}}, \quad J\in \mathbb R^+.
\end{align}
Note that the sum over $i_1,\cdots, i_q$ is over all $N^q$ sequences with repetitions allowed which is different from fermionic SYK.
As dissipators in the Lindbladian, let us consider
\begin{align}
    L_i = \sqrt{\gamma}x_i,
    \quad 
    i = 1,\cdots, N,
\end{align}
where $\gamma$ controls the strength of dissipation.
We could also consider 
$
    L_i = \sqrt{\gamma}p_i,
    $
    $
    i = 1,\cdots, N
    $
and many other possibilities. 
With the Hamiltonian and dissipators, the evolution of the density matrix $\rho(t)$ is given by
\begin{align}
\frac{d\rho}{dt} = -i [H, \rho]
+
\sum_i \left[L_i \rho L^{\dag}_i
-\frac{1}{2} \{L^{\dag}_i L_i, \rho\}
\right].
\end{align}

We should note that, unlike the fermionic counterparts, the spectrum of the bosonic SYK Hamiltonian \eqref{bosonic SYK}
may not be bounded both from the above and below.
When it is not bounded from below, the system is unstable.
The simplest case is obtained by switching off $J$, and taking $v$ to be negative,
i.e., 
the inverted harmonic oscillator 
potential.
Even when $v>0$, for a particular realization of $J_{i_1\cdots i_q}$, the system may be unstable. 
For unstable cases, the system may not reach a stationary state, at least in the absence of dissipation.
Nevertheless, in the following, we will add Lindbladian 
dissipation, in which case the system may reach stationarity,
even when the Hamiltonian part is unstable. 
Having this in mind, in the following, we consider both signs of $v$.
Unbounded potentials 
and their interplay with Lindbladian dissipation has also been studied in 
\cite{Cho2025}
in matrix quantum mechanics.

We now invoke the operator-state map or the vectorization to 
represent the Lindbladian as an  operator 
acting on state kets
(mapped from density operators) 
in the doubled Hilbert space,
${\cal H}\otimes {\cal H}^*
={\cal H}_+ \otimes {\cal H}_-
$.
We denote operators 
acting on ${\cal H}_+$
as 
${O}$ or ${O}^+$,
and 
the corresponding operators
acting on ${\cal H}_-$
as
$\tilde{{O}}$ or 
${O}^-$.
With this procedure, 
the Lindbladian 
acting on ${\cal H}_+\otimes {\cal H}_-$ is represented as
\begin{align}
    &{\cal L}
    = -i H^+ + i 
    (H^{-})^T
    \nonumber \\
    & 
    +
    \sum_i \left(
     L_{i}^+ (L_{i}^-)^* 
    -\frac{1}{2}
     (L^+_{i})^{\dag} L^{+}_{i}
    -\frac{1}{2}
     (L_{i}^-)^T (L^-_{i})^*
\right).
\end{align}
Explicitly, the Lindbladian is given by
\begin{align}
    {\cal L}
&=
- i \sum_i \frac{p_i^2}{2m}
+ i \sum_i \frac{\tilde{p}_i^2}{2m}
    -i v \sum_i x_i^2 +i v \sum_i \tilde{x}_i^2
\nonumber \\
&\quad
    -i \sum_{i_1,\cdots, i_q} J_{i_1\cdots i_q}x_{i_1} \cdots x_{i_q}
    +
    i \sum_{i_1,\cdots, i_q} J_{i_1\cdots i_q}\tilde{x}_{i_1} \cdots \tilde{x}_{i_q}
    \nonumber \\
    &\quad
    +\gamma \sum_i 
    \left(
        x_i \tilde{x}_i
        - 
        \frac{1}{2} x_i x_i
        -\frac{1}{2} \tilde{x}_i \tilde{x}_i
        \right).
\end{align}

By vectorization, the master equation is represented in the doubled Hilbert space as
$
d|\rho(t)\rangle/d t = {\cal L}|\rho(t)\rangle.
$
In terms of an initial state $\rho_0$, the solution to the master equation 
can be written as
$|\rho(t) \rangle 
=
e^{t {\cal L}}|\rho_0\rangle
\equiv
(\rho(t)\otimes I) |I\rangle$,
where 
$|\rho_0\rangle$
and $|I\rangle$
are the mapped states from the 
initial density operator $\rho_0$
and the identity operator (the infinite temperature Gibbs state), respectively.
%
%
We note, by construction, the identity state
is a fixed point of the master equation,
${\cal L}|I\rangle=\langle I|{\cal L}=0$.
The physical observables,
e.g.,
the expectation value of 
an operator $O$, 
are 
represented as
\begin{align}
\mathrm{Tr}_{{\cal H}}\, [O \rho(t)]
=
\langle I| O^+ |\rho(t)\rangle,
\quad 
O^+ = O\otimes I.
\end{align}
This can alternatively be represented 
by using the Heisenberg operator,
\begin{align}
O^+(t):= e^{-t {\cal L}}\, O^+\, e^{t {\cal L}},
\end{align}
as 
$\langle I| O^+(t)|\rho_0\rangle$.
In the following, we are mostly interested in
the Green's function,
\begin{align}
G_{ab}(t,t') =
\frac{1}{N}\sum_i
\langle I| T (x^a_i(t) x^b_i(t'))|\rho_0\rangle,
\end{align}
where $x^a_i(t)$ is the Heisenberg operator $x^a_i$ ($a=\pm 1$),
$x^a_i(t) = e^{-t {\cal L} } x^a_i e^{t{\cal L}}$,
and $T$ represents time-ordering.

\subsection{Symmetries of the Lindbladian}
\label{Symmetries of the Lindbladian}

Let us briefly discuss the symmetry properties of the Lindbladian.
One of its most fundamental symmetries 
is the modular conjugation symmetry
\cite{Kawabata_2023c},
expressed as
${\cal J} {\cal L}{\cal J}^{-1}={\cal L}$,
where
the antiunitary modular conjugation operator ${\cal J}$
acting on
$x_i, \tilde{x}_i$ and $p_i, \tilde{p}_i$
as
\begin{align}
&
{\cal J} x_i {\cal J}^{-1} = \tilde{x}_i,
\quad
{\cal J} \tilde{x}_i {\cal J}^{-1} = {x}_i,
\nonumber \\
&
{\cal J} p_i {\cal J}^{-1} = \tilde{p}_i,
\quad
{\cal J} \tilde{p}_i {\cal J}^{-1} = {p}_i,
\nonumber \\
&
{\cal J} z {\cal J}^{-1} = z^*,
\quad 
z\in \mathbb{C}.
\end{align}
The modular conjugation symmetry ensures that the density operator remains hermitian during time evolution,
$\rho(t)^{\dag}=\rho(t)$.
In terms of the vectorized state $|\rho(t)\rangle$,
${\cal J}|\rho(t)\rangle = |\rho(t) \rangle$.
We also note ${\cal J}|I\rangle = |I\rangle$.

The modular conjugation symmetry implies symmetries of observables.  
As we are interested in the stationary properties,
we assume $|\rho(t)\rangle$
converges to a stationary state in the long-time limit $t\to \infty$,
$|\rho(t)\rangle \to |\rho_{s}\rangle$,
where $|\rho_s\rangle$ is a right eigenvector of ${\cal L}$.
The real part of the corresponding eigenvalue is zero, i.e., an infinite lifetime. We are then interested in expectation values (correlation functions) of the form 
$\langle I| O |\rho_s\rangle
=\mathrm{Tr}_{{\cal H}} [ \rho_s O]
$.
To proceed further, we have to specify additional details of $\rho_s$.
In the following, 
we assume $\rho_s=I$
and explore the consequences. 
(In fact, the following analysis remains valid as long as we use the 
identical left- and right-eigen kets.)

Under this assumption, we can easily see
that correlation functions satisfy
\begin{align}
&
\big\langle O \big\rangle = 
\big\langle {\cal J}O{\cal J} \big\rangle^*,
\\
&
\big\langle O^a O^{\prime b} \big\rangle
=
\big\langle O^{\prime b} O^a \big\rangle,
\end{align}
where $\langle \cdots \rangle = \langle I|\cdots |I\rangle$,
${}^*$ denotes complex conjugation, 
$O$ is an arbitrary operator acting on 
${\cal H}_+\otimes {\cal H}_-$,
and $O^{a}$ and $O^{\prime b}$
are arbitrary operators 
acting 
on ${\cal H}_{a}$ and ${\cal H}_b$, respectively,
with $a,b=\pm 1$.
In particular, 
for the position operators $x^a_i$
and $x^b_j$, we find
\begin{align}
&
\big\langle x^a_i x^b_j \big\rangle = 
\big\langle x^{\bar{a}}_i x^{\bar{b}}_j\rangle^*,
\\
&
\big\langle x^a_i x^b_j \big\rangle
=
\big\langle x^b_j x^a_i \big\rangle,
\end{align}
where $\bar{a}=-a$.
The latter is an avatar of 
the Kubo-Martin-Schwinger (KMS) relation 
at infinite temperature.
For the Heisenberg operators,
similar symmetry relations
hold: 
\begin{align}
&
\big\langle O(t)O'(t')\cdots \big\rangle = 
\big\langle {\cal J}O(t)O'(t')\cdots{\cal J} \big\rangle^*,
 \\
&
\big\langle O^a(t) O^{\prime b}(t') \big\rangle
=
\big\langle 
[O^{\prime b \dagger}(t')]^{\dag}
[O^{a \dag}(t)]^{\dag} \big\rangle.
\end{align}
Here, 
$O^{a \dag}(t)$
is the Heisenberg operator of $O^{a\dag}$,
$
O^{a \dag}(t) \equiv e^{-t {\cal L}}O^{a \dag} e^{+t {\cal L}}
$, which is in general not equal to
$[O^{a}(t)]^{\dag}$.
In particular,
for the position operators,
noting ${\cal J} {\cal L}{\cal J}={\cal L}$,
we obtain
\begin{align}
&
\big\langle x^a_i(t) x^b_j(t') \big\rangle = 
\big\langle x^{\bar{a}}_i(t) x^{\bar{b}}_j(t')\big\rangle^*,
\label{sym 1}
\\
&
\big\langle x^a_i(t) x^b_j(t') \big\rangle
=
\big\langle 
[x^b_j(t')]^{\dag}
[x^a_i(t)]^{\dag} \big\rangle.
\label{sym 2}
\end{align}
Once again,
Eq.\ \eqref{sym 2} is nothing but the Kubo-Martin-Schwinger relation 
at infinite temperature;
when 
$
x^{a \dag}(t) \equiv e^{-t {\cal L}}x^{a \dag} e^{+t {\cal L}}
=
[x^a(t)]^{\dag}
$,
Eq.\ \eqref{sym 2} reduces to 
$
\langle x^a_i(t) x^b_j(t') 
\rangle
=
\langle 
x^b_j(t')
x^a_i(t) \rangle
$,
which is the KMS relation in a more familiar form.

In the next section, as we formulate the Schwinger-Keldysh path integral approach, we will find that similar symmetries are realized in the Schwinger-Dyson equation in the large $N$ limit.

\begin{widetext}
\subsection{Schwinger-Keldysh path integral}
The physical quantities of
interest, 
such as the 
dissipative form factor
$
Z = 
\mathrm{Tr}_{{\cal H}_+ \otimes {\cal H}_-}
\left[
e^{t \mathcal{L}}
\right]
$
or correlation functions
$
\mathrm{Tr}_{{\cal H}_+ \otimes {\cal H}_-}
\left[
{\cal O}_1
{\cal O}_2 
\cdots 
e^{t \mathcal{L}}
\right]
$
can be expressed 
using the 
coherent state path 
integral in the doubled space,
or 
the Schwinger-Keldysh path integral
\cite{Sieberer_2016}.
Explicitly, it is
given by
\begin{align}
\label{starting action}
    Z &= \int \mathcal{D}J\, P[J]\, \int {\cal D}x \int {\cal D}\tilde{x}
    \exp\Bigg[
    \int dt\,\Big\{ 
    +i  \sum_i \frac{m}{2} (\dot{x}_i)^2
    -i  \sum_i \frac{m}{2} (\dot{\tilde{x}}_i)^2
    \nonumber \\
    & \quad 
    -i  \sum_{i_1,\cdots, i_q} J_{i_1\cdots i_q}x_{i_1} \cdots x_{i_q}
    +
    i  \sum_{i_1,\cdots, i_q} J_{i_1\cdots i_q}\tilde{x}_{i_1} \cdots \tilde{x}_{i_q}
    \nonumber \\
    &\quad 
    -i v \sum_i x_i^2
    +i v \sum_i \tilde{x}_i^2
    +\gamma \sum_i 
    \left(
        x_i \tilde{x}_i
        - 
        \frac{1}{2} x_i x_i
        -\frac{1}{2} \tilde{x}_i \tilde{x}_i
        \right)
        \Big\}
    \Bigg]
\end{align}
We can now integrate over disorder
and then study the resulting path integral in the large $N$ limit.
Intermediate steps of the analysis are presented in
Appendix 
\ref{Schwinger-Keldysh path integral and large N analysis}.
To this end, we introduce 
a bilocal collective field, 
\begin{align}
    \label{eq:collective-field}
    G_{ab}(t_1,t_2) = \frac{1}{N}
    \sum_i x^a_i(t_1)x^b_i(t_2)
\end{align}
together with 
another auxiliary field 
$\Sigma(t_1,t_2)$.
Integrating over $x^a$ and $\tilde{x}^a$,
the Schwinger-Keldysh 
action (Lindbladian) is 
given by
\begin{align}
    \label{eq:collective-field-action}
    \frac{\mathcal{L}}{N}
    =&-\frac{1}{2}
    \text{Tr}\log (D)
    -\int dt_1 dt_2
    \sum_{a,b=+,-}
    \left\{
        \frac{J^2}{4} 
        s_{ab}G_{ab}(t_1,t_2)^q
    \right.
    \nonumber\\
    &\left.
        +i \tilde\Sigma_{ab}(t_1,t_2) 
        G_{ab}(t_1,t_2)
        - \delta(t_1-t_2)
        \left(
            \begin{array}{cc}
                -\frac{\gamma}{2}-iv & \frac{\gamma}{2} \\
                \frac{\gamma}{2} & - \frac{\gamma}{2}+iv
            \end{array}
        \right)_{ab} 
        G_{ab}(t_1,t_2)
    \right\},
\end{align}
where $s_{ab} = +1$ if $a=b$ and $s_{ab} = -1$ if $a\neq b$.
\end{widetext}From the Schwinger-Keldysh action, 
taking the derivative of $\tilde{\Sigma}$
and $G$, we obtain the Schwinger-Dyson equations,
\begin{align}
\label{eq:kadanoff-baym}
    &G(t_1,t_2)
    ={-\frac{1}{2}}D^{-1}(t_1,t_2),
    \nonumber \\ 
    &\tilde{\Sigma}_{ab}(t_1,t_2)
    =
    i\frac{J^2q}{4}s_{ab}G_{ab}^{q-1}(t_1,t_2)
\nonumber \\
&\qquad - i\delta(t_1-t_2)
        \left(
            \begin{array}{cc}
                -\frac{\gamma}{2}-iv & \frac{\gamma}{2} \\
                \frac{\gamma}{2} & - \frac{\gamma}{2}+iv
            \end{array}
        \right)_{ab}, 
\end{align}
where 
\begin{align}
    \label{eq:define_D}
    D(t_1, t_2) = \frac{i m}{2} 
    \left(
  \begin{array}{cc} 
   1 & 0 \\
   0 & -1
  \end{array}
  \right)
    \delta(t_1-t_2)
    \partial_t^2
    + i\tilde\Sigma(t_1, t_2).
\end{align}

At late times (both $t_1, t_2$ are large), if a stationary state is realized,
the Green's functions and
self-energies depend only on the difference $t=t_1-t_2$.
This allows us to write the Schwinger-Dyson equations as
\begin{align}\label{SD_equations}
    G(\omega)
    &={-\frac{1}{2}}D^{-1}(\omega),
    \nonumber \\ 
    \tilde{\Sigma}_{ab}(t)
    &=
    i\frac{J^2q}{4}s_{ab}G_{ab}^{q-1}(t)
    - i\delta(t)
        \left(
            \begin{array}{cc}
                -\frac{\gamma}{2}-iv & \frac{\gamma}{2} \\
                \frac{\gamma}{2} & - \frac{\gamma}{2}+iv
            \end{array}
      \right)_{ab}.
\end{align}
It is also useful to study 
the on-shell action (the value of the action at the saddle point),
which is obtained by substituting the saddle point equations into Eq.\ \eqref{eq:collective-field-action},
\begin{align}
    \label{eq:saddle-point-action}
    \frac{\mathcal S_{\text{sp}}}{N}
    &=-\frac{1}{2}
    \int \frac{d\omega}{2\pi}\log (D(\omega))
    \nonumber \\
    &\quad
    +\int dt
    \sum_{a,b=+,-}
    \left\{
        \frac{J^2}{4} 
        (q-1)s_{ab}G_{ab}(t)^q
    \right\}.
\end{align}

Let us now examine the symmetries of the saddle point equations. 
First, consider the transformation $G_{ab}(t_1, t_2) \rightarrow G_{\bar a \bar b}(t_1, t_2)^*$. 
The second equation in Eq.\ \eqref{eq:kadanoff-baym} prompts us to transform the self-energy as $\Sigma_{ab}(t_1, t_2) \rightarrow -\Sigma_{\bar a \bar b}(t_1, t_2)^*$. Then, from Eq.\ \eqref{eq:define_D} we have $D_{ab}(t_1, t_2) \rightarrow D_{\bar a \bar b}(t_1, t_2)^*$ which is indeed consistent with the first equation in Eq.\ \eqref{eq:kadanoff-baym}. Thus, these three transformations taken together also leave the saddle point equations invariant. 
Second, consider the transformation $G_{ab}(t_1,t_2) \rightarrow
G_{ba}(t_2, t_1)$.
Correspondingly, we transform the self-energy and the
kernel $D$ as 
$\Sigma_{ab}(t_1, t_2) \rightarrow \Sigma_{ba}(t_1,t_2)$  and
$D_{ab}(t_1,t_2) \rightarrow D_{ba}(t_1,t_2)$. 
These together leave the saddle-point equations \eqref{eq:kadanoff-baym} invariant.
We have thus established that the saddle-point equations possess both of these symmetries. 
Focusing on stationary solutions and on the Green's functions, these symmetries are summarized as 
\begin{align}
\label{eq: conjugation}
G_{ab}(t) &= G_{\bar{a}\bar{b}}(t)^*,
\\
\label{eq: KMS}
G_{ab}(t) &= G_{ba}(-t).
\end{align}
Following the discussion in Sec.\ \ref{Symmetries of the Lindbladian} using the operator formalism, we refer to these two symmetries 
of the saddle-point equations as modular conjugation symmetry and
the KMS relation,
as they correspond to
Eqs.\ \eqref{sym 1}
and \eqref{sym 2}.
\footnote{
Although we call Eqs.\ \eqref{sym 1} and \eqref{sym 2} as symmetries of the saddle point equations, the KMS relation here should not be confused with the KMS symmetry
\cite{2015PhRvB..92m4307S,Liu_2021}
which requires an additional antiunitary symmetry (such as time-reversal).}

While we draw an analogy between the symmetries in the operator formalism and those in the saddle-point equations, the precise correspondence is somewhat unclear. 
First, the saddle-point equations are derived in the large‑$N$ limit, whereas the operator formalism remains valid for arbitrary $N$.
Moreover, in the operator formalism, the identification of symmetries in expectation values relies on the assumption that the left and right eigenvectors are identical. This assumption is not explicitly encoded in the saddle-point equations; rather, a particular stationary solution is dynamically selected, which may break the equivalence between the left and right eigenvectors.

%

In the next section, these symmetries will play a crucial role in our characterization of saddle-point solutions and in identifying the dynamical phase diagram. We will see that some solutions to the saddle-point equations spontaneously break these symmetries. When such symmetry-breaking saddles become dominant, this signals a phase transition in the system.

\begin{figure*}[t]
    \centering
    \includegraphics[scale=0.8]{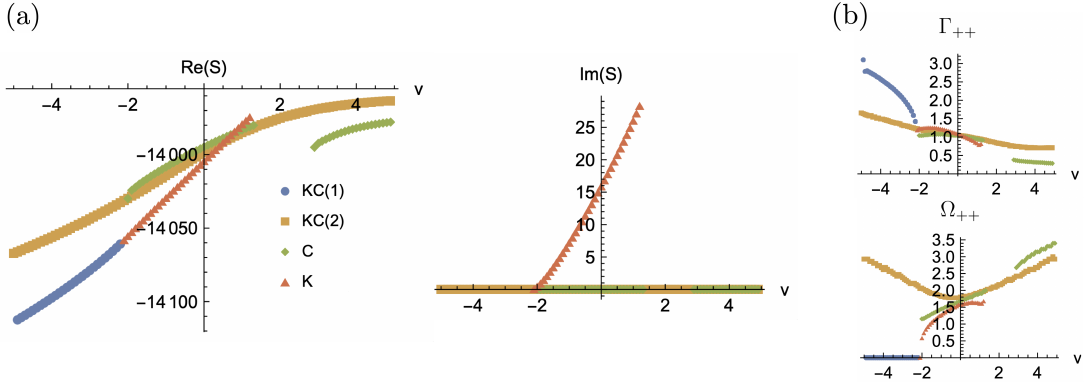}
    \caption{
    \label{fig:action}
    (a)
   Numerical values of the on-shell action for each solution along $J=5, \gamma=4$ for $v=-5$ to $v=5$. The different solutions 
   are labeled by their symmetries: "K" represents the KMS relation and "C" denotes modular conjugation symmetry. 
   "KC(1)" and "KC(2)" are two distinct solutions that preserve both symmetries. 
  (b)
   Decay rate ($\Gamma_{++}$)
   and frequency ($\Omega_{++}$)
   extracted from $G_{++}(\omega)$
   at $J=5, \gamma=4$ for $v=-5$ to $v=5$.  
   }
\end{figure*}

\section{Saddle point solutions}

\subsection{The case when $J=0$}


Let us begin by discussing the behaviors of the model when $J=0$, which is amenable to analytical treatment. 
It is convenient to use 
the Heisenberg equations of motion, 
$dO/dt = [O, {\cal L}]$, 
\begin{align}
\dot{x} &= p/m,
\quad 
\dot{p} = -2 v x - i \gamma (\tilde{x}-x),
\nonumber \\
\dot{\tilde{x}} &= -\tilde{p}/m,
\quad
\dot{\tilde{p}} = +2 v x + i \gamma (\tilde{x}-x).
\end{align}
(As degrees of freedom with different $i$ decouple, we suppress the index $i$.)
Equivalently,
the coordinates $x$ and $\tilde{x}$ obey
the second-order differential equation,
\begin{align}
&
D_0=
    \left(
        \begin{array}{cc}
            -\frac{im}{2}\partial_{t}^2
            -iv
            -\frac{\gamma}{2} & \frac{\gamma}{2} \\
            \frac{\gamma}{2} & \frac{im}{2}\partial_{t}^2+iv-\frac{\gamma}{2}
        \end{array}
    \right) ,
    \quad 
    \nonumber \\\
    &
    D_0
\left(
\begin{array}{c}
x\\
\tilde{x}
\end{array}
\right)
=0.
\label{heis eq}
\end{align}
In terms of the linear combinations
$X_{\pm}\equiv x\pm \tilde{x}$ and $P_{\pm}\equiv p\pm \tilde{p}$, the Heisenberg equation of motion can be rewritten as
\begin{align}
\dot{X}_+ &= P_-/m,
\quad 
\dot{P}_- =
-2v X_+ - 2 i\gamma X_-
\nonumber\\ 
\dot{X}_- &= P_+/m,
\quad 
\dot{P}_+ = - 2 v X_-.
\end{align}

We see that the time-dependence of 
the linear combination $X_-$
is not affected by 
the dissipation $\gamma$,
and the solution is simply given by a 
linear superposition of
$\cos( t \sqrt{2v/m})$
and
$\sin( t \sqrt{2v/m})$.
On the other hand,
$X_+(t)$
depends linearly on $\gamma$,
and is given as a linear superposition 
of
$t (\gamma/v) \cos (\sqrt{2v/m} t)$,
$t (\gamma/v) \sin (\sqrt{2v/m} t)$,
and 
$(\gamma/v) \sin (\sqrt{2v/m} t)$.
However, it does not exhibit exponential 
decay $\sim e^{- C \gamma t}$
by dissipation (where $C$ is some constant).
These observations indicate
that our dissipation is not sufficiently "strong" or "mixing".
We will however see that once we include the interaction effect $J\neq 0$, we do find that the cases where the Green functions
decay exponentially in time.
In contrast, for the fermionic model, nonzero dissipation induces 
exponential decay, and when $J\neq 0$, the zero-dissipation limit was of interest due to its anomalous diffusion behavior.


On the other hand,
when negative, the $v$-term serves as a dissipation,
since the inverted harmonic potential 
renders the system effectively open
and the spectrum is continuous.
In the coming subsections, we will discuss the competition
between $\gamma$ and the interaction $J$,
and also $v$ and $J$.
We also note that the case $v=0$, which separates the open and closed systems, is somewhat special. In this case, $X_-$ is a first-order polynomial in $t$, while $X_+$ is third order.

Finally, we record the explicit form of the Green functions. 
The saddle point equation 
when $J=0$
is given by
\begin{align}
&
D_0
    \left(
        \begin{array}{cc}
            G_{++}(t,t') & G_{+-}(t,t') \\
            G_{-+}(t,t') & G_{--}(t,t')
        \end{array}
    \right) = \frac{i}{2}\left(
        \begin{array}{cc}
            1 & 0 \\
            0 & 1
        \end{array}
    \right) \delta(t-t').
\end{align}
The solution is given by
\begin{align}
    G_{ab}(t) &= \frac{\sqrt{\pi } \text{sgn}(t)}{8 (m v)^{3/2}} \left[- \gamma  t \sqrt{\frac{2v}{m}} \cos \left( t \sqrt{\frac{2v}{m}}\right) \right. 
    \nonumber 
    \\
    &\quad 
    \left. +  (\gamma -2(a+b) i v) \sin \left( t \sqrt{\frac{2v}{m}}\right)\right].
\end{align}

\subsection{Numerical solutions for $J\neq 0$}

\begin{figure}[t]
    \centering
    \includegraphics[scale=0.5]{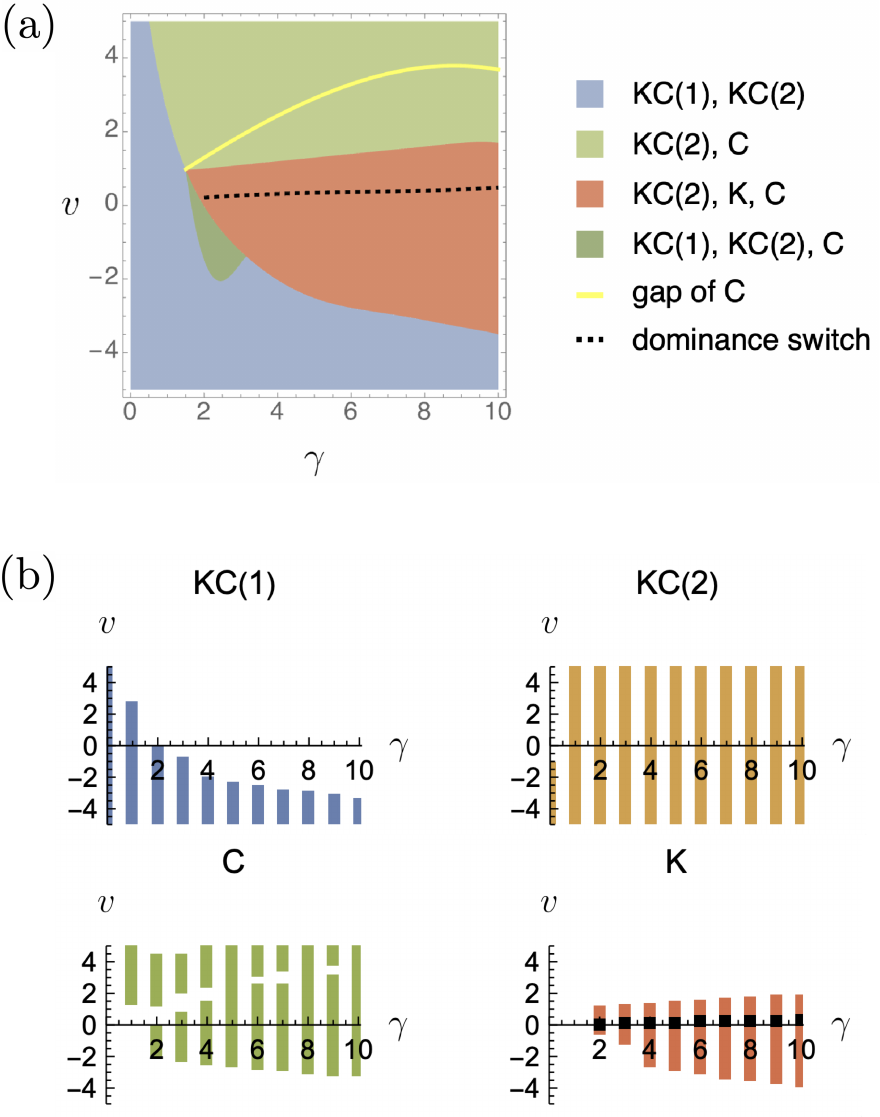}
    \caption{
    \label{fig:phasediag}
    (a) Putative phase diagram for $J=5$   
    based on the spontaneous breaking 
    symmetries.
    (b) Various types of saddle-point solutions for $J=5$ 
    used to generate the phase diagram in (a).
    Here, the colored data points indicate the existence of the solutions.
    In the final “K” plot, the black points indicate where the dominance between solution types switches. 
    }
\end{figure}

We now turn to the case with nonzero SYK interaction $(J \neq 0)$, and focus on the quartic case $q = 4$ throughout. We set $m=1$ for simplicity.
In the presence of SYK interactions,
the Schwinger–Dyson equations must be solved numerically.
To study stationary-state properties, we focus on time-translationally invariant solutions,
$G_{ab}(t, t’) = G_{ab}(t - t’)$ and $\tilde{\Sigma}_{ab}(t, t’) = \tilde{\Sigma}_{ab}(t - t’)$.
In our numerical implementation, we impose periodic boundary conditions in $t - t’$ with sufficiently large period.
We filter out non-stationary solutions by examining whether the solution vanishes at late times. Specifically, we integrate the numerical values of the Green's function solution in the late-time domain; the resulting value indicates how closely the solution converges to 0 in this region. Solutions are discarded if this integrated value exceeds a predefined threshold;
see below for the typical threshold values used in our calculations.

Depending on the choice of the parameters
$J, \gamma, v$, multiple stationary solutions may exist. 
We employ two different strategies to obtain distinct solutions. The first strategy begins with a random initial ansatz at fixed $\gamma$ and 
large enough $|v|$, $v = \pm 5$, which converges to a solution. We proceed by scanning along the $v$-axis, using the previously obtained solution as the initialization for the next point. This scan is performed in both directions of $v$. The second strategy involves explicitly enforcing symmetries in the numerical procedure. Starting from a random initialization, we impose symmetry constraints during each iteration. The two symmetries considered are the KMS relation \eqref{eq: KMS}
and modular conjugation symmetry \eqref{eq: conjugation}.


In Figs.\ \ref{Green's functions a}, \ref{Green's functions b}
and \ref{Green's functions c}
in Appendix
\ref{Green's functions in various phases}, we present representative numerical solutions obtained at fixed parameters $J=5$ and $\gamma=4$, using three different values of $v$ corresponding to the three major phases characterized by distinct symmetries. 
At $v=-4$, we find two solutions, both preserving the KMS symmetry and modular conjugation symmetry. 
At $v=1$, we identify three solutions, each exhibiting different symmetries: one preserving only the KMS relation, 
one preserving only modular conjugation symmetry, and 
a third one preserving both symmetries.
At $v=3$, we again observe two solutions: one preserving only modular conjugation symmetry and the other preserving both symmetries. We label these solutions by the symmetries they preserve:
\begin{itemize}
  \item KC(1): A first solution that preserves both the KMS symmetry and modular conjugation symmetry. It appears in the region of $v<-2$ for $\gamma=4$ and $J=5$, as shown in Fig.\ \ref{fig:action}. 
  \item KC(2): A second solution that also preserves both symmetries. It exists across the full range of $v$ values for $\gamma=4$ and $J=5$.
  \item K: A solution preserving only the KMS relation.
  \item C: A solution preserving only modular conjugation symmetry.
\end{itemize}
We note that the convergence of solutions at late times changes noticeably around $v=1.3$, as detected by the numerical integration over a small window in the late-time regime. For $v < 1.3$, the integral remains below $10^{-10}$, indicating good convergence towards zero. However, for $v>1.3$, the integral increases to the order of $0.1$, particularly for the "C" solutions.


For a precise physical interpretation of the symmetry breaking, we recall that the derivation of the symmetry relations \eqref{sym 1} and \eqref{sym 2} in the operator formalism assumes that $\rho_s$ coincides with the identity operator $I$
(or more generally, that the left and right eigenvectors of ${\cal L}$ are identical when computing expectation values). 
We expect that, in the symmetry-breaking solutions, it is this equivalence between the left and right eigenvectors that breaks down—rather than a failure of the modular conjugation symmetry of $\rho_s$. The latter would imply a violation of hermiticity, which would render the saddle-point solutions unphysical.

When multiple solutions exist, we need a criterion to determine which solution is dominant. In Hermitian systems and in Euclidean path integrals, dominance is typically determined by comparing the on-shell values of the action. We may adopt a similar approach here and use the action in Eq.\ \eqref{eq:collective-field-action} (or Eq.\ \eqref{eq:saddle-point-action}) to evaluate the relative dominance of solutions.
In the Schwinger–Keldysh path integral, however, the on-shell action can in general be complex. We indeed find that this is the case for the “K” solution, which breaks modular conjugation symmetry. Although complex on-shell actions are also allowed in the fermionic SYK Lindbladian model, such behavior was not observed there~\cite{Kawabata_2023b}.
The appearance of a complex action and the associated symmetry breaking is a distinctive feature of the bosonic model.

In Fig.\ \ref{fig:action}(a), 
we plot the action for $J = 5$, $\gamma = 4$, varying $v$ from $-5$ to $5$. In the intermediate region, the real parts of the actions intersect near $v \approx 0.5$, indicating a change in the dominant solution.
We also note that
the "C" solutions and "KC(2)" solutions in the region of $v\in(-2,1.3)$ have actions that are nearly identical. 
As shown in Fig.\ \ref{Green's functions b}, at $v=1$, the corresponding solutions are very similar, except that the "KC(2)" solution has a real-valued $G_{+-}$, while "C" solution exhibits a nonzero imaginary component in $G_{+-}$. 


Our findings are summarized in Fig.\ \ref{fig:phasediag} as a putative phase diagram of the model. 
Recall that, in the absence of the SYK interaction, the system exhibits two distinct “phases” for $v > 0$ and $v < 0$, characterized respectively by oscillatory and decaying behavior of the Green’s functions. 
When the interaction is introduced, new phases and phase transitions emerge due to the competition between dissipation and interaction effects.

In Fig.\ \ref{fig:phasediag}, we identify the green and blue regions as remnants of the $v > 0$ and $v < 0$ phases from the non-interacting limit. 
We note that the blue region, which corresponds to the $v < 0$ phase in the non-interacting case, is deformed by the interaction and now extends into the $v > 0$ domain when $\gamma$ is sufficiently small.
This deformation bears resemblance to the anomalous diffusion observed in the fermionic SYK Lindbladian: in the non-interacting model, the decay rate vanishes as $v \to 0^-$, while in the presence of interactions, it remains finite even as $v \to 0$.

For intermediate values of $v$ and sufficiently large $\gamma$, we observe a new region (colored in orange) that arises from the interplay between dissipation and interactions. 
Furthermore, within this region, the dominant saddle-point solution changes as $v$ is varied (marked by black dots), signaling an additional phase transition
inside the orange region.

Additional features also emerge. First, within the green region, there appears to be a narrow transition zone where the branch of the “C” solution either becomes discontinuous or vanishes entirely 
-- See the region $1.5 < v<3$ in Fig.\ \ref{fig:action}(a).
Second, we find a small dark-green region between the orange and blue phases, where three distinct solutions coexist.

To further characterize the behavior of the Green’s functions across different solutions, we fit the numerical data using the ansatz
$G_{ab}(t)=A_{ab}\exp(-\Gamma_{ab} t)\cos(\Omega_{ab} t)$. 
To implement this numerically, we first compute the Fourier transform of the solutions. The corresponding frequency-domain Green's function is typically a broad distribution over $\omega$, rather than a simple function with a few poles. Therefore, extracting a peak value directly from the spectrum as $\Omega$ would be inaccurate. This Fourier transform analysis is used only to distinguish oscillatory solutions and non-oscillatory ones. For non-oscillation solutions, the frequency-domain Green's function $G_\omega$ is peaked at $\omega=0$, allowing us to exclude them from further fitting, since an infinite period is numerically difficult to resolve within a finite time window. Next, we identify the local maxima of the Green functions in the time domain and obtain $\Gamma$ by taking the logarithm of their amplitudes and doing a linear fit. To accurately determine the oscillation frequency, we multiply the original solution by $\exp(\Gamma t)$, so that the effect of decay is removed. This step is crucial: Without it, the positions of local maxima would shift slightly, leading to an offset in the estimated frequency.  
The extracted frequencies and decay rates
for $G_{++}$ are shown in Fig.\ \ref{fig:action}(b). 
The behaviors for the off-diagonal Greens' function, namlely $\Gamma_{+-}$ and $\Omega_{+-}$,
are very similar and not shown.

\section{Discussion/summary}

In this work, we have studied a class of bosonic SYK-like models with Lindbladian dissipation. Unlike their fermionic counterparts, these bosonic models exhibit distinctive features arising from marginal stability and strong correlations. Using the large-$N$ technique, we analyzed the collective dynamics and the behavior of two-point functions, 
uncovering various phases
with different numbers of saddle point solutions. 
Despite potential instabilities associated with 
the inverted potential, 
our analysis shows that a well-defined saddle point exists in the large-$N$ limit.

Our results suggest that the bosonic SYK Lindbladians can serve as a useful theoretical laboratory for exploring strongly coupled dissipative systems. While we have not computed the Lyapunov exponent or other diagnostics of chaos in this work, doing so remains an important direction for future research
\cite{garciagarcia2024lyapunovexponentsignaturedissipative}. 
In particular, it would be interesting to investigate whether these models exhibit chaotic behavior and potentially saturate the chaos bound, as in the fermionic SYK model 
\cite{ garcíagarcía2024lyapunovexponentsignaturedissipative}.
For the fermionic variants of our model,  the Lyapunov exponent was found to vary continuously with the coupling to the bath, eventually becoming negative at a critical value, signaling a transition to a dynamics which is no longer quantum chaotic.

Another promising direction is to explore connections to the $p$-spin glass model, which also involves the spherical constraint among bosonic degrees of freedom. Extending our analysis to such models may provide further insights into the landscape of solvable systems and their phase structure.
Additional future directions include the application of various techniques developed in,
e.g., 
\cite{cho2025bootstrappingnonequilibriumstochasticprocesses,
Cho2025,
nishimura2025quantumdecoherencecaldeiraleggettmodel},
and also
investigating potential holographic duals.
Finally, it would also be worthwhile to examine possible experimental realizations in platforms such as cold atoms or programmable quantum simulators
\cite{baumgartner2024quantumsimulationsachdevyekitaevmodel}.

\begin{acknowledgments}
We thank Minjae Cho and Tokiro Numasawa for helpful discussions. We are particularly grateful to Minjae Cho for sharing Ref.\ \cite{Cho2025} prior to its arXiv submission 
and for kindly agreeing to coordinate the timing of our submissions. This work was supported by the National Science Foundation under Grant No.\ DMR-2409412.
\end{acknowledgments}

\appendix

\begin{widetext}
\section{Schwinger-Keldysh path integral and large $N$ analysis}

\label{Schwinger-Keldysh path integral and large N analysis}

In this appendix, we provide additional details of the large $N$ analysis.
Our starting point is the 
Schwinger-Keldysh path integral
\eqref{starting action}.
Integrating over disorder,
we obtain
\begin{align}
    Z &= 
    \int {\cal D}x \int {\cal D}\tilde{x}
    \exp\Bigg[
    \int dt\,\Big\{ 
    +i  \sum_i \frac{m}{2} (\dot{x}_i)^2
    -i  \sum_i \frac{m}{2} (\dot{\tilde{x}}_i)^2
    -i v \sum_i x_i^2
    +i v \sum_i \tilde{x}_i^2
    \nonumber \\
    &\quad 
    +\gamma \sum_i 
    \left(
        x_i \tilde{x}_i
        - 
        \frac{1}{2} x_i x_i
        -\frac{1}{2} \tilde{x}_i \tilde{x}_i
        \right)
        \Big\}
        -\frac{J^2}{4N^{q-1}} \int dt_1 dt_2
        \sum_{a,b=+,-} s_{ab}\left(
        \sum_i x_i^a(t_1) x_i^b(t_2)
        \right)^q
    \Bigg],
\end{align}
where $s_{ab} = +1$ if $a=b$ and $s_{ab} = -1$ if $a\neq b$.
Using the bilocal collective field $G_{ab}(t_1, t_2)$ in \eqref{eq:collective-field} and 
an auxiliary field $\Sigma(t_1,t_2)$, the Schwinger-Keldysh path integral 
can be rewritten as 
\begin{align}
    Z &= 
    \int \mathcal{D}G\mathcal{D}[\Sigma]
    \int {\cal D}x \int {\cal D}\tilde{x}
    \exp\Bigg[
    \int dt\,\Big\{ 
    +i  \sum_i \frac{m}{2} (\dot{x}_i)^2
    -i  \sum_i \frac{m}{2} (\dot{\tilde{x}}_i)^2
    -i v \sum_i x_i^2
    +i v \sum_i \tilde{x}_i^2
 \nonumber \\
 &\quad 
    +\gamma \sum_i 
    \left(
        x_i \tilde{x}_i
        - 
        \frac{1}{2} x_i x_i
        -\frac{1}{2} \tilde{x}_i \tilde{x}_i
        \right)
        \Big\}
        \nonumber \\
        &\quad 
        -
        \int dt_1 dt_2
        \sum_{a,b=+,-}
        \left\{
        \frac{J^2 N}{4} s_{ab}G_{ab}(t_1,t_2)^q
        +i N \Sigma_{ab}(t_1,t_2) 
        \Big(
        G_{ab}(t_1,t_2)
        - (1/N)\sum_i x_i^a(t_1)x_i^b(t_2)
        \Big)
        \right\}
    \Bigg].
\end{align}
We can integrate over 
$x^a_i$
and $\tilde{x}^a_i$,
leading to
\begin{align}
    Z &= 
    \int \mathcal{D}G\mathcal{D}[\Sigma]
    \, 
    \mathrm{Det}^{-N/2}\left[D\right]
    \exp\Bigg[
        -
        \int dt_1 dt_2
        \sum_{a,b=+,-}
        \left\{
        \frac{J^2 N}{4} s_{ab}G_{ab}(t_1,t_2)^q
        +i N \Sigma_{ab}(t_1,t_2) 
        G_{ab}(t_1,t_2)
        \right\}
    \Bigg],
\end{align}
where the operator $D$
is given by 
\begin{align}
    D(t_1, t_2) = \frac{i m}{2} 
    \left(
  \begin{array}{cc} 
   1 & 0 \\
   0 & -1
  \end{array}
  \right)
    \delta(t_1-t_2)
    \partial_{t_2}^2
    +
    \delta(t_1-t_2)
    \left(
        \begin{array}{cc}
            -\gamma/2 -iv& \gamma/2 \\
            \gamma/2 & - \gamma/2+iv
        \end{array}
        \right)
    + i\Sigma(t_1, t_2).
\end{align}
Here, the matrix entries are ordered using the $(+, -)$ convention.

We can shift $\Sigma$ to simplify the saddle point equations as
\begin{align}
    \tilde\Sigma(t_1, t_2)=
    \Sigma(t_1, t_2)
    -i \delta(t_1-t_2)
    \left(
        \begin{array}{cc}
            -\frac{\gamma}{2}-iv & \frac{\gamma}{2} \\
            \frac{\gamma}{2} & - \frac{\gamma}{2}+iv
        \end{array}
    \right).
\end{align}
Correspondingly, 
$D$ can be written as \eqref{eq:define_D}.
These manipulations lead to
the Schwinger-Keldysh 
action (Lindbladian)  
\eqref{eq:collective-field-action}
and 
the Schwinger-Dyson equation
\eqref{eq:kadanoff-baym}.

\section{Green's functions in various phases}
\label{Green's functions in various phases}

In this appendix, we present the numerical solutions of Green's functions for $J=5$ and $\gamma=4$ with varying  $v=-4, 1$, and $3$. Each value of $v$ corresponds to a distinct phase shown in Fig.~\ref{fig:action}.

\begin{figure*}[h]
    \centering
    \includegraphics[width=0.75\textwidth]{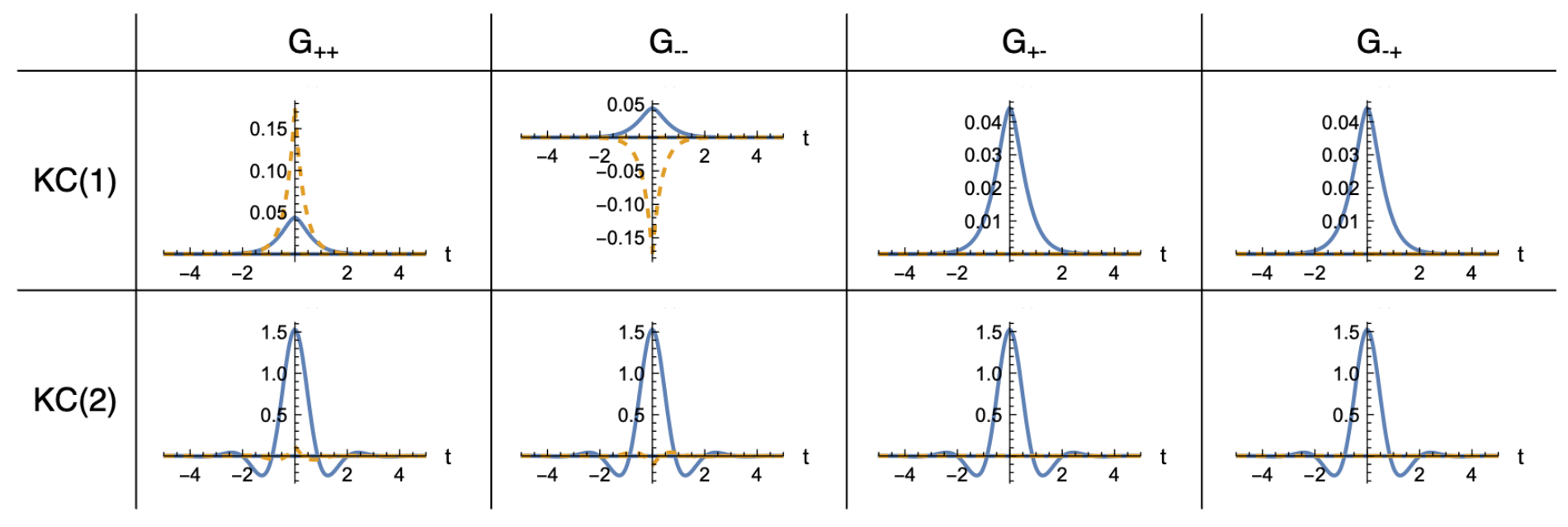}
    \caption{
   \label{Green's functions a}
   Two distinct solutions for $J=5, \gamma=4, v=-4$.
   The solid (dashed) lines represent the real (imaginary) part of the solutions. The notation "KC" denotes solutions that preserve both KMS and Conjugation symmetries. To distinguish between them, we refer to them "KC(1)" and "KC(2)", consistent with the notation used in Fig.\ \ref{fig:action}. The numerical calculations are performed over the time domain $[-25,25]$, which is truncated to $[-5,5]$ for a better presentation. The following figures are similarly truncated for clarity.}
    
\end{figure*}

\begin{figure*}[h]
    \centering
    \includegraphics[width=0.75\textwidth]{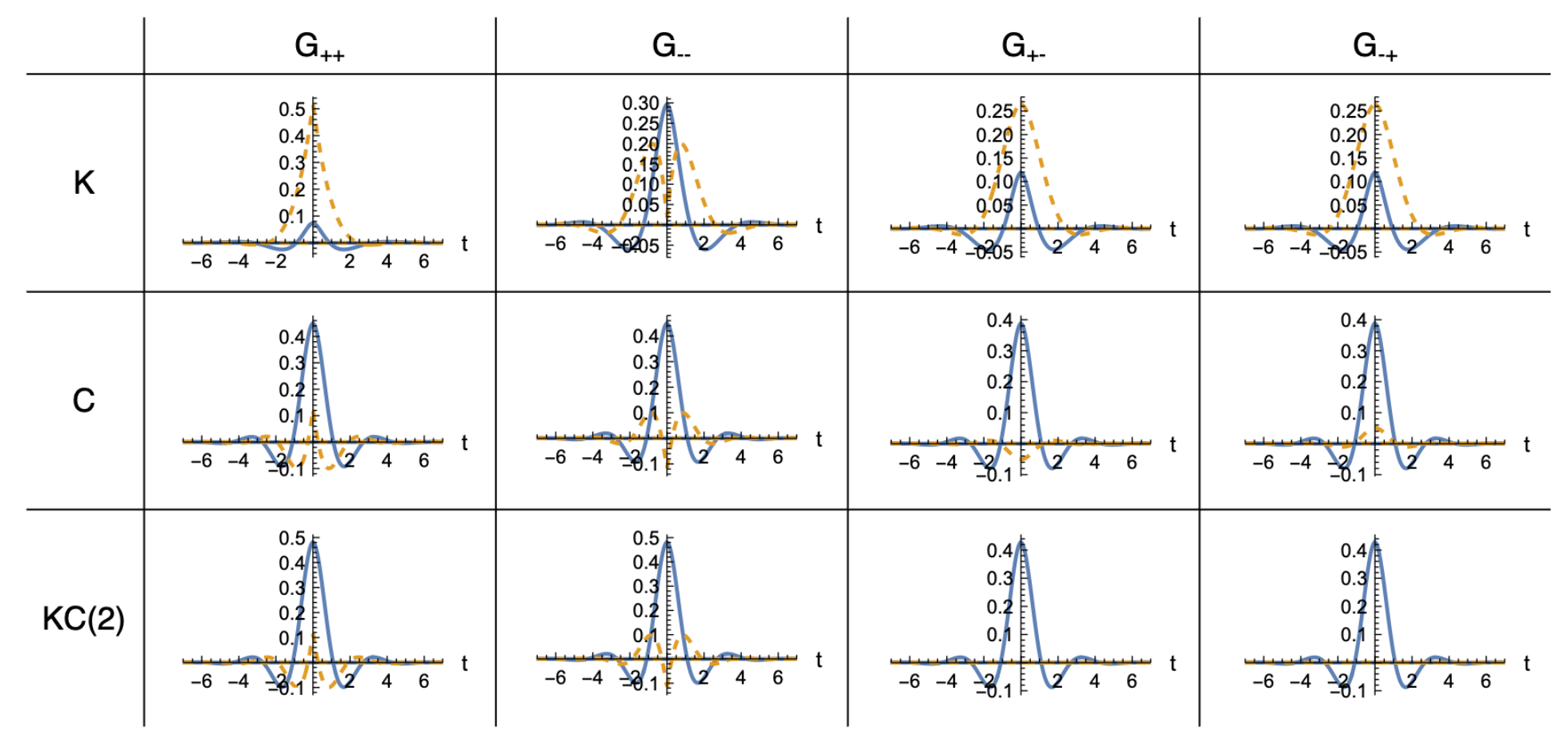}
    \caption{
   \label{Green's functions b}
   Three Solutions for $J=5, \gamma=4, v=1$. The label "K" denotes solutions that preserve only the KMS relation; modular conjugation symmetry is broken. "C" indicates that only modular conjugation symmetry is preserved while the KMS relation is broken. "KC(2)" solution preserves both KMS and conjugation symmetries. 
    }
\end{figure*}

\begin{figure*}[h]
    \centering
    \includegraphics[width=0.75\textwidth]{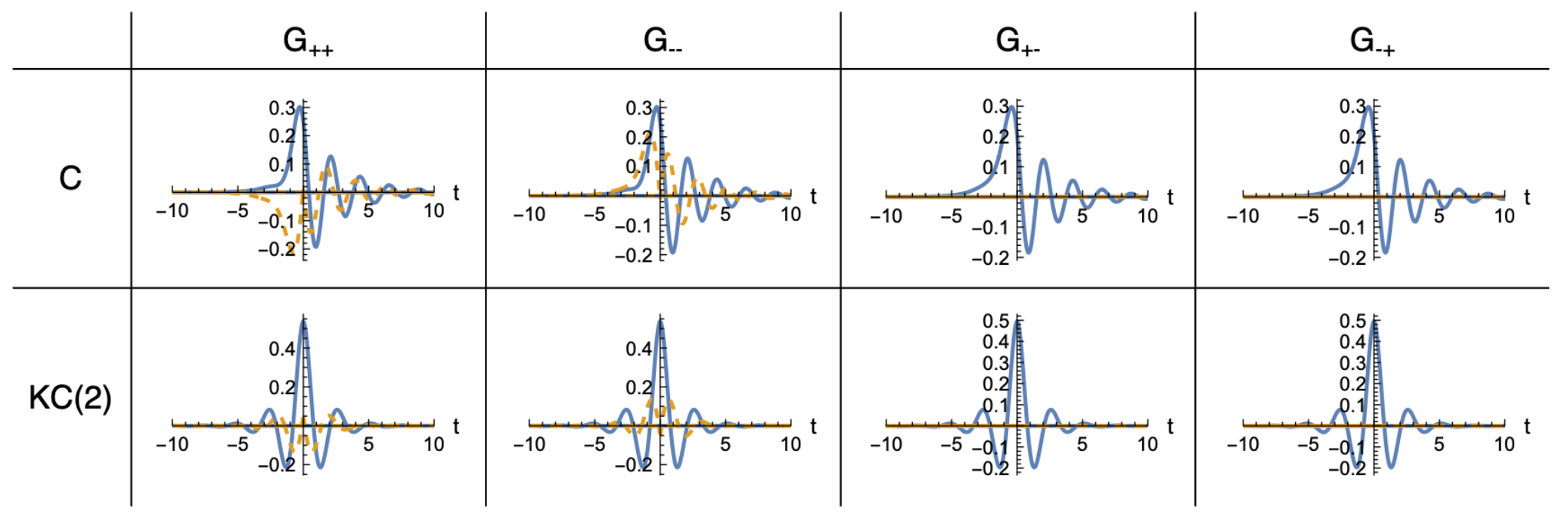}
    \caption{
   \label{Green's functions c}
   Two solutions for $J=5, \gamma=4, v=3$. The label "C" represents a solution that breaks KMS while preserving modular conjugation symmetry. Unlike other stationary solutions, this solution shows a slower convergence at late times. The "KC(2)" solution preserves both KMS and Conjugation symmetries. 
    }
\end{figure*}

\end{widetext}
\bibliography{ref}
\end{document}